\documentclass[a4paper,12pt]{article}

\usepackage[usenames]{color}

\usepackage{graphicx}

\usepackage{array}

\usepackage{amsmath}
\usepackage{amssymb}
\usepackage{theorem}
\usepackage{epsf}

\theoremstyle{break}

\def\therefore{.\raise1ex\hbox{.}.}
\def\because{\raise1ex\hbox{.}.\raise1ex\hbox{.}}

\title{\bf{Lorentz invariant and supersymmetric interpretation of noncommutative quantum field theory}}

\author{Yoshishige Kobayashi \thanks{E-mail : yosh@phys.metro-u.ac.jp}
\\
Shin Sasaki \thanks{E-mail : shin-s@phys.metro-u.ac.jp}}

\date{\empty}

\begin{document}
\maketitle
\begin{center}
{\it Department of Physics, Faculty of Science\\
     Tokyo Metropolitan University\\
     1--1 Minami-osawa, Hachioji-shi\\
     Tokyo 192--0397, Japan}
\end{center}

\vspace{2cm}

\begin{abstract}
In this paper, using a Hopf-algebraic method,
we construct deformed Poincar\'e SUSY algebra in terms 
of twisted (Hopf) algebra. By adapting this twist deformed super-Poincar\'e algrebra as our fundamental symmetry, 
we can see the consistency between the algebra and non(anti)commutative
 relation among (super)coordinates and interpret that symmetry of
 non(anti)commutative QFT is in fact twisted one. The key point is validity of our 
new twist element that guarantees non(anti)commutativity of space. It is 
checked in this paper for $\mathcal{N} = 1$ case. We also comment on the 
possibility of noncommutative central charge coordinate. Finally,
 because our twist operation does not break the original algebra, we can
 claim that (twisted) SUSY is not broken in contrast to the string
 inspired $\mathcal{N}=1/2$ SUSY in $\mathcal{N}=1 $ non(anti)commutative superspace.

\end{abstract}

\section{Introduction}
Quantum field theory on the noncommutative space-time has been under intensive study in recent years. 
This idea originates from Snyder's work \cite{Snyder}. They considered 
space-time noncommutativity can smear the UV divergence and give naive cut  
off of the theory. 
Stringy origin of noncommutativity of 
space-time is realized by considering the string propagating on {\it 
constant} NS-NS B-field background \cite{NC_brane}. In that situation, we can find the space 
coordinates on D-brane become noncommutative, {\it i.e.}
\begin{equation}
[x^i,x^j]=i \Theta^{ij} \not=0,
\end{equation}
by taking appropriate zero-slope limit. 
Here, noncommutative parameter $\Theta^{ij}$ naively corresponds to 
VEV of background field $B_{\mu \nu}$. This {\it constant} parameter 
thus breaks ordinary Lorentz invariance. 

On the other hand, the $\mathcal{N}=1$ non(anti)commutative superspace has also 
been studied \cite{Seiberg,Ooguri_Vafa}. This can be realized when there are R-R 
background (constant self-dual graviphoton backgrond) in the superstring 
propagating space. In that case, 
non(anti)commutativity of $\mathcal{N}=1$ superspace is realised on 
world sheet boundary
\begin{equation}
\{\theta^{\alpha},\theta^{\beta} \} = C^{\alpha\beta} \not=0 .
\end{equation}
Here, $C^{\alpha\beta}$ is naively VEV of constant graviphoton field.

Quantum field theory on these non(anti)commutative (super)space has been 
studied in recent decade. For example, see the review 
\cite{Szabo,Douglas} and references therein for QFT on noncommutative 
space-time and recent work \cite{Araki_Ito,Ferrara_Lledo,Klemm_Penati,Ooguri_Vafa,Berkovits_Seiberg,Boer_Grassi} for QFT on non(anti)commutative superspace. There are many interesting properties 
in the framework of NC QFT {\it e.g.} UV/IR mixing, noncommutative instanton, 
relation to matrix model and so on. But almost all these works have been discussed in a 
formally Lorentz invariant approach and their representation corresponds 
to usual Poincar\'e algebra in spite of their violation of Lorentz symmetry.

But recently, deformed Lorentz invariance of this 
NC QFT was proposed \cite{Nishijima,Chaichian,Oeckl}  by using twisted Hopf 
algebra. By reinterpreting our fundamental symmetry as twisted deformed
Hopf algebra, we can construct twisted Lorentz invariant quantum field theory on noncommutative space.

In this paper, we extend twisted deformed (Hopf) Poincar\'e algebra to 
twisted deformed {\it Poincar\'e SUSY} algebra and investigate whether this 
approach would be to what extent extensible. By choosing appropriate twist element 
$\mathcal{F} \in \mathcal{U(SP)} \otimes \mathcal{U(SP)}$, we see
twisted Lorentz invariant formulation of QFT on {\it non(anti)commutative
superspace} is possible. And we also comment on the new noncommutativity of 
central charge coordinate.

The organization of this paper is as follows. Section \ref{Hopf_algebra}
is a brief review of deformed Poincar\'e algebra especially twisted 
equation of $P$-$P$ twist element and the work of
Ref.\cite{Nishijima}. Section \ref{SUSY} is extention to simple SUSY algebra
and we see validity of our new fermionic twist element. Consistency
between algebra and twisted symmetry is discussed in section
\ref{consistency}. In section 
\ref{extended},  we consider extended ($\mathcal{N} \ge 2$) SUSY algebra and
establish some other kind of twist element. Then we claim new noncommutativity of central
charge coordinate. Section \ref{Conclusion} is our conclusion.

\section{Twisted deformed (Hopf) Poincar\'e algebra \label{Hopf_algebra}}
In this section, we review the twisted deformed Hopf Poincar\'e algebra 
following the recent work \cite{Nishijima}. The review of the Hopf algebra itself is in 
Ref.\cite{Hopf,guide_to_quantum_group,quantum_group}. 

The Poincar\'e algebra $\mathcal{P}$ consists of Lorentz generators $M_{\mu \nu}$ and 
translation generators $P_{\mu}$. This algebra contains abelian subalgebra 
$P_{\mu}$. By using this subalgebra, we can construct twist 
element of quantum group. For more detail see Ref.\cite{guide_to_quantum_group}. 
This twist element permits to deform the universal enveloping of the 
Poincar\'e algebra $\mathcal{U(P)}$ (this is called trivial Hopf algebra
that possesses quasitriangularity).
There exist co-product $\Delta_0 : \mathcal{U(P)} \longrightarrow 
\mathcal{U(P)} \otimes \mathcal{U(P)}$. For $X \in \mathcal{P}$, this is 
written as 
\begin{equation}
\Delta_0 (X) = X \otimes \mathbf{1} + \mathbf{1} \otimes X,
\end{equation}
and our co-unit $\epsilon: \mathcal{U(P)}\longrightarrow \mathcal{K}$
 and antipode $\gamma: \mathcal{U(P)}\longrightarrow \mathcal{U(P)}$
are
\begin{eqnarray}
\epsilon(X) &=& 0 \ (\mathrm{co \ unit}) ,\nonumber \\
\gamma(X) &=& -X \ (\mathrm{antipode}),
\end{eqnarray}
for all $X \in \mathcal{P}$ and
\begin{eqnarray}
\epsilon({\mathbf{1}}) &=& 1, \nonumber \\
\gamma(\mathbf{1}) &=& \mathbf{1}.
\end{eqnarray}
$\mathcal{K}$ is the base field of the vector space.
After twisting the algebra by twist element $ \mathcal{F}$, the 
co-product of twisted algebra $\mathcal{U}_t \mathcal(P)$ is redefined by 
$ \Delta_t (X) = \mathcal{F} \Delta_0 (X) \mathcal{F}^{-1}$. The 
consistency of Hopf algebra requires that after twisting, we have to 
change the definition of multiplication of original algebra 
representation \cite{guide_to_quantum_group,Kulish_Mudrov} 
\begin{eqnarray}
m ( a \otimes b) \equiv ab \longrightarrow m_t (a \otimes b) &\equiv& a 
\star b \nonumber \\
&=& m \circ \mathcal{F}^{-1} ( a \otimes b).
\end{eqnarray}
This formally can be considered as the origin of noncommutativity of space.

Twist element is constructed from abelian subalgebra 
$P_{\mu}$ \cite{Reshetikhin}
\begin{equation}
\mathcal{F}^{PP} = \exp \left( \frac{i}{2} \Theta^{\mu \nu} P_{\mu} \otimes 
P_{\nu} \right) \label{P_twist}.
\end{equation}
To be consistent with the property 
of Hopf algebra, the twist element $\mathcal{F}$ should satisfy twist equation
\begin{equation}
\mathcal{F}_{12} ( \Delta_0 \otimes \mathrm{id}) (\mathcal{F}) = 
\mathcal{F}_{23} ( \mathrm{id} \otimes \Delta_0) (\mathcal{F}) \label{twist_eq},
\end{equation}
and co-unit condition
\begin{eqnarray}
(\epsilon \otimes \mathrm{id})(\mathcal{F}) = \mathbf{1} = ( \mathrm{id} 
\otimes \epsilon)(\mathcal{F}).
\end{eqnarray}
Notice that the co-product $\Delta_0$ property admit
\begin{eqnarray}
 \left( \mathrm{id} \otimes \Delta_0 \right) e^{X \otimes Y} &=& \left( \mathrm{id} \otimes \Delta_0 \right) \sum^{\infty}_{n=0} \frac{1}{n !} \left( X \otimes Y \right)^n \nonumber \\
&=& \left( \mathrm{id} \otimes \Delta_0 \right) \sum_n \frac{1}{n !} X^n \otimes Y^n \nonumber \\
&=& \sum_n \frac{1}{n!} X^n \otimes \Delta_0 \left( Y ^n \right) \nonumber \\
&=& \sum_n \frac{1}{n!} X^n \otimes \left( \Delta_0 (Y)  \right)^n \nonumber \\
&=& \sum_n \frac{1}{n!} \left( X \otimes \Delta_0 (Y) \right)^n = e^{X \otimes \Delta_0 (Y)}
\end{eqnarray}
for bosonic quantities $X,Y$. By using this property and bosonic feature of generator $P_{\mu}$, we 
can see this $P$-$P$ twist element acturally satisfy the twist equation
as follows.
The $P$-$P$ twist element is the sum of the infinite series:
\begin{eqnarray}
\mathcal{F} &=& \exp\left(\frac{i}{2} \Theta^{\mu\nu}P_\mu\otimes P_\nu \right) \nonumber \\
	&=& \sum_{n=0}^{\infty} \frac{1}{n!} \Bigl( \frac{i}{2} \Bigr) ^n
	\Theta^{\mu_1 \nu_1} \cdots \Theta^{\mu_n \nu_n}
	(P_{\mu_1} \otimes P_{\nu_1}) \cdots (P_{\mu_n} \otimes P_{\nu_n}) \nonumber \\
	&=&  \sum_{n=0}^{\infty} \frac{1}{n!} \Bigl( \frac{i}{2} \Bigr) ^n
	\Theta^{\mu_1 \nu_1} \cdots \Theta^{\mu_n \nu_n}
	(P_{\mu_1}\cdots P_{\mu_n}) \otimes (P_{\nu_1}\cdots P_{\nu_n}) \ .
\end{eqnarray}
Co-product acts on the product $P_\mu$s in such a way that
\begin{eqnarray}
\Delta_0( P_{\mu_1} \cdots P_{\mu_n} ) 
	&=& \prod_{i=1}^{n}
	 ( P_{\mu_i} \otimes \mathbf{1} 
	  + \mathbf{1} \otimes  P_{\mu_i} ) \nonumber \\
	&=& \sum_{l=0}^n \binom{n}{l} P^l \otimes P^{n-l} \ ,
\end{eqnarray}
here $P^l$ stands for the $l$ times product of $P_\mu$ abstractly.
Using above, and after appropriate reassignment of indices, 
we get
\begin{eqnarray}
\lefteqn{
 \mathcal{F}_{12} ( \Delta_0 \otimes \mathrm{id}) (\mathcal{F})
	  = 
	 \exp\left( \frac{i}{2} \Theta^{\rho \sigma} P_\rho \otimes P_\sigma \right)_{12} 
	( \Delta_0 \otimes \mathrm{id})
	\exp\left( \frac{i}{2} \Theta^{\mu \nu} P_\mu \otimes P_\nu \right) }\nonumber  \\
	& = & 
	\sum_{n=0}^\infty \sum_{m=0}^\infty 
	\sum_{l=0}^n  
	\frac{1}{n!m!} \Bigl( \frac{i}{2} \Bigr) ^{n+m}
	 \binom{n}{l}
	\Theta^{\rho_1 \sigma_1} \cdots \Theta^{\rho_m \sigma_m}
	\Theta^{\mu_1 \nu_1} \cdots \Theta^{\mu_n \nu_n} \nonumber \\
	& & \times P_{\rho_1} \cdots P_{\rho_m}  P_{\mu_1} \cdots P_{\mu_l}
	\otimes
 	P_{\sigma_1} \cdots P_{\sigma_m}  P_{\mu_{l+1}} \cdots P_{\mu_n}
	\otimes
	 P_{\nu_{1}} \cdots P_{\nu_n}  \ , 
\end{eqnarray}
and
\begin{eqnarray}
	\lefteqn{\mathcal{F}_{23} ( \mathrm{id} \otimes \Delta_0) (\mathcal{F})}
	\nonumber \\
	& = &
	\sum_{n'=0}^\infty \sum_{m'=0}^\infty
	\sum_{l'=0}^{n'}
	\frac{1}{n'!m'!} \Bigl( \frac{i}{2} \Bigr) ^{n'+m'} 
	\binom{n'}{l'} 
	\Theta^{\rho_1 \sigma_1} \cdots \Theta^{\rho_{m'} \sigma_{m'}}
	\Theta^{\mu_1 \nu_1} \cdots \Theta^{\mu_{n'} \nu_{n'}} \nonumber \\
	& &  \times P_{\mu_1} \cdots P_{\mu_{n'}}
	\otimes
	P_{\rho_1} \cdots P_{\rho_{m'}} P_{\nu_1} \cdots P_{\nu_{l'}}
	\otimes
 	P_{\sigma_1} \cdots P_{\sigma_m'} P_{\nu_{l'+1}} \cdots P_{\nu_n} \ 
	. 
\end{eqnarray}
Each individual term in expansion can be characterized only by a set of three numbers;
 namely $\alpha$, $\beta$ and $\gamma$.
$\alpha$ represents the number of $P_\mu P_\nu$ contraction
through $\Theta$ between the first and second factors of the tensor product, 
$\beta$ between the first and third factors,
$\gamma$ between the second and third factors.
And we see the relation
\begin{equation}
\left\{
\begin{array}{cccc}
\alpha = & m & = & l' \nonumber \\
\beta = & l & = & n' - l' \nonumber \\
\gamma = & n-l& = & m'
\end{array}
\right. ,
\end{equation}
then $n,m,l$ and $n',m',l'$ are determined from these equations.
If the term which relates to $\alpha$, $\beta$ , $\gamma$ is found 
in $n,m,l$ series, 
corresponing term always exist in $n',m',l'$ expansion, 
and vice versa. 
Moreover these coefficients are identical,
\begin{equation}
\frac{1}{m!n!}
\Bigl( \frac{i}{2} \Bigr) ^{n+m}
\binom{n}{l}
=
\frac{1}{m'!n'!}
\Bigl( \frac{i}{2} \Bigr) ^{n'+m'}
\binom{n'}{l'}
=
\frac{1}{\alpha!  \beta! \gamma!}
\Bigl( \frac{i}{2} \Bigr) ^{\alpha + \beta + \gamma} \ .
\end{equation}
So the twist equation is satisfied order by order. 

Co-unit condition is rather trivial because of the property
\begin{equation}
\epsilon(XY) = \epsilon(X)\epsilon(Y).
\end{equation}

If we consider this $P$-$P$ twist element and deform the
original algebra, the multiplication $m$ of original algebra 
representation has to be
changed in the deformed algebra $\mathcal{U}_t\mathcal{(P)}$,
\begin{eqnarray}
m(x_{\mu} \otimes x_{\nu}) = x_{\mu} x_{\nu} \longrightarrow m_t (x_{\mu} \otimes x_{\nu}) = x_{\mu} \star x_{\nu} 
\end{eqnarray}
This suggests noncommutativity of space-time coordinate. Indeed, because 
translation generator acts on coordinate as $P_{\mu} x_{\nu} =  i \eta_{\mu 
\nu}$, we find
\begin{eqnarray}
m_t (x_{\mu} \otimes x_{\nu}) &=& x_{\mu} \star x_{\nu} = m \circ e^{ - \frac{i}{2} \Theta^{\alpha \beta} P_{\alpha} \otimes P_{\beta}} ( x_{\mu} \otimes x_{\nu}) \nonumber \\
&=& m \circ \left[ x_{\mu} \otimes x_{\nu} + \frac{i}{2} \Theta^{\alpha \beta} \eta_{\alpha \mu} \otimes \eta_{\beta \nu} \right] \nonumber \\
&=& x_{\mu} x_{\nu} + \frac{i}{2} \Theta_{\mu \nu},
\end{eqnarray}
then this gives noncommutative relation among space-time coordinates
\begin{equation}
[x_{\mu},x_{\nu}]_{\star} = i \Theta_{\mu \nu} \not= 0.
\end{equation}
This exactly correspond to Moyal-Weyl star product bracket.\footnote{In practice, we
are considering space-space noncommutativity because time direction
noncommutativity may alarm the causality and unitarity problem.}

\section{Construction of twisted Poincar\'e SUSY algebra \label{SUSY}}
It is not difficult to extend ordinary Hopf algebra to super
($\mathbf{Z}_2$-graded) Hopf algebra \cite{super_Hopf}. 
The definition of Hopf algebra basically contains graded algebra. So, we 
can include fermionic generator, {\it i.e.} supercharge $Q_{\alpha}$. The key point is how to construct the twist element $\mathcal{F} \in 
\mathcal{U(SP)} \otimes \mathcal{U(SP)}$. Here $\mathcal{U(SP)}$ is 
universal enveloping of Poincar\'e SUSY algebra. We first consider the simplest $\mathcal{N}=1$ SUSY algebra
\begin{eqnarray}
[P_{\mu},Q_{\alpha}] &=& 0 , \quad [M_{\mu \nu} , Q_{\alpha}] = i ( 
\sigma_{\mu \nu})_{\alpha}^{\;\, \beta} Q_{\beta}, \nonumber \\
\{Q_{\alpha}, \bar{Q}_{\dot{\beta}}\} &=& 2 \sigma^{\mu}_{\alpha 
\dot{\beta}} P_{\mu} , \quad \{ Q_{\alpha}, Q_{\beta} \} = 0.
\end{eqnarray}
We here omit $\bar{Q}_{\dot{\alpha}}$ sector. Because Hopf algebra is 
defined for this graded algebra, we can choose abelian subsector as $P_{\mu}, 
Q_{\alpha}$ sector. Choosing these generators lead to new twist 
element in $\mathcal{U(SP)} \otimes \mathcal{U(SP)}$, for example
\begin{equation}
\mathcal{F}^{QQ} = \exp \left[ - \frac{1}{2} C^{\alpha \beta} Q_{\alpha} 
\otimes Q_{\beta} \right], \label{twist_Q}
\end{equation}
$C^{\alpha \beta}$ in the equation is some constant. This twist element
satisfies the Yang-Baxter equation
\begin{equation}
\mathcal{F}_{12} \mathcal{F}_{13} \mathcal{F}_{23} = \mathcal{F}_{23} \mathcal{F}_{13} \mathcal{F}_{12}.
\end{equation}
Because supercharge $Q_{\alpha}$ is grassmann odd, the exponential does 
end at finite order
\begin{eqnarray}
\mathcal{F} &=& \exp\left[ - \frac{1}{2} C^{\alpha \beta} Q_{\alpha} 
\otimes Q_{\beta} \right] \nonumber \\
&=& \mathbf{1} \otimes \mathbf{1} - \frac{1}{2} C^{\alpha \beta} 
Q_{\alpha} \otimes Q_{\beta} - \frac{1}{8} C^{\alpha \beta} C^{{\alpha}' 
{\beta}'} Q_{\alpha} Q_{{\alpha}'} \otimes Q_{\beta} Q_{{\beta}'}.
\end{eqnarray}

It is not obvious that this element satisfy the twist
equation because of their non-trivial sign that arise when interchanging
the two grassmann odd quantities. But this finite expansion of exponent
enable us to check the validity of this $Q$-$Q$ twist by straightforward calculation. Let us check whether this twist element satisfy the twist equation
(\ref{twist_eq}). The 
left hand side of Eq.~(\ref{twist_eq}) is
\begin{eqnarray}
\mathrm{LHS} &=& \mathcal{F}_{12}( \Delta_0 \otimes 
\mathrm{id})(\mathcal{F}) \nonumber \\
&=& \left( \mathbf{1} \otimes \mathbf{1} - \frac{1}{2} C^{\alpha \beta} 
Q_{\alpha} \otimes Q_{\beta} - \frac{1}{8} C^{\alpha \beta} C^{\gamma 
\delta} Q_{\alpha } Q_{\gamma} \otimes Q_{\beta} Q_{\delta} \right)_{12} 
\nonumber \\
& & \times \left[ \Delta_0 (\mathbf{1}) \otimes \mathbf{1} - \frac{1}{2} 
C^{\kappa \tau } \Delta_0 (Q_{\kappa}) \otimes Q_{\tau} - \frac{1}{8} C^{\kappa \tau} C^{\lambda \pi} \Delta_0 
(Q_{\kappa} Q_{\lambda}) \otimes Q_{\tau} Q_{\pi} \right] \nonumber
\end{eqnarray}
\begin{eqnarray}
&=& \left( \mathbf{1} \otimes \mathbf{1} - \frac{1}{2} C^{\alpha \beta} 
Q_{\alpha} \otimes Q_{\beta} - \frac{1}{8} C^{\alpha \beta} C^{\gamma 
\delta} Q_{\alpha } Q_{\gamma} \otimes Q_{\beta} Q_{\delta} \right)_{12} 
\nonumber \\
& &  \times \left[ \mathbf{1} \otimes \mathbf{1} \otimes \mathbf{1} - 
	    \frac{1}{2} C^{\kappa \tau} \left( Q_{\kappa} \otimes 
	    \mathbf{1} \otimes Q_{\tau} + \mathbf{1} \otimes Q_{\kappa} 
	    \otimes Q_{\tau}  \right) \right. \nonumber \\
& &  \quad - \frac{1}{8} C^{\kappa \tau} C^{\lambda \pi} \left( Q_{\kappa 
} Q_{\lambda} \otimes \mathbf{1} \otimes Q_{\tau} Q_{\pi} + Q_{\kappa} 
\otimes Q_{\lambda} \otimes Q_{\tau} Q_{\pi}  \right. \nonumber \\
& &  \qquad \qquad \quad \quad \quad - \left. \left. Q_{\lambda} \otimes Q_{\kappa} \otimes Q_{\tau} 
Q_{\pi} + \mathbf{1} \otimes Q_{\kappa} Q_{\lambda} \otimes Q_{\tau} 
Q_{\pi} \right) \right] \label{LHS}.
\end{eqnarray}
and similar for RHS. We used here the graded tensor product property
\begin{equation}
\left(a \otimes b \right) \left( a' \otimes b' \right) = (-)^{|b| |a'|} 
\left( a a' \otimes b b' \right),
\end{equation}
where $|a|$ is fermion number of $a$, and also co-product property
\begin{eqnarray}
\Delta_0 (XY) &=& \Delta_0 (X) \Delta_0 (Y) \nonumber \\
&=& \left( X \otimes \mathbf{1} + \mathbf{1} \otimes X \right) \left( Y 
\otimes \mathbf{1} + \mathbf{1} \otimes Y \right) \nonumber \\
&=& XY \otimes \mathbf{1} + X \otimes Y + (-)^{|X||Y|}Y \otimes X + \mathbf{1} 
\otimes YX.
\end{eqnarray}
for $ X,Y \in \mathcal{SP}$. Be careful, after expanding these 
quantities explicitly, for higher order $\mathcal{O}(C^3)$ term that 
involve three supercharge in one tensor sector gives vanishing 
contribution. Then, it is not difficult to confirm that left hand side 
is  precisely equal to right hand side. So the twist element quantity Eq.~(\ref{twist_Q}) is valid. 
This twist leads to non(anti)commutativity of $\mathcal{N}=1$ superspace 
coordinate. Actually, because supercharge acts on fermionic coordinate 
$\theta^{\alpha}$ as $Q_{\alpha} \theta^{\beta} = i 
\delta_{\alpha}^{\beta}$, we see
\begin{eqnarray}
m_t ( \theta^{\alpha} \otimes \theta^{\beta}) &=& \theta^{\alpha} 
\star \theta^{\beta} \nonumber \\
&=& m \circ e^{ \frac{1}{2} C^{\gamma \delta} Q_{\gamma} \otimes 
Q_{\delta}} ( \theta^{\alpha} \otimes \theta^{\beta}) \nonumber \\
&=& m \circ \left[ \theta^{\alpha} \otimes \theta^{\beta} + \frac{1}{2} 
C^{\gamma \delta} \delta_{\gamma}^{\alpha} \otimes 
\delta_{\delta}^{\beta} \right] \nonumber \\
&=& \theta^{\alpha} \theta^{\beta} + \frac{1}{2} C^{\alpha \beta}.
\end{eqnarray}
This gives 
\begin{equation}
\{ \theta^{\alpha} , \theta^{\beta }\}_{\star} = C^{\alpha \beta} \not=0.
\end{equation}
Simultaneously, addtional effects occur on other commutaton relations,
\begin{eqnarray}
& &\left[x^{\mu},x^{\nu} \right]_{\star} = C^{\alpha \beta} 
\sigma^{\mu}_{\alpha \dot{\gamma}} \sigma^{\nu}_{\beta \dot{\delta}} 
\bar{\theta}^{\dot{\gamma}} \bar{\theta}^{\dot{\delta}}, \nonumber \\
& &\left[x^{\mu} , \theta^{\alpha} \right]_{\star} = - i C^{\alpha 
\beta} \sigma^{\mu}_{\beta \dot{\gamma}} \bar{\theta}^{\dot{\gamma}},
\end{eqnarray}
it may be eliminated if one takes chiral coordinates. 

This corresponds to $Q$-deformation (non-supersymmetric deformation) of 
non(anti)commutative supersymmetric theory \cite{ferrara2}\footnote{But
in twisted SUSY case, half of SUSY does not broken because our
original algebra is intact after twisting.}. 

Note, however, we can choose only
chiral part {\it or} anti-chiral part of supercharge because there exist
non-trivial anticommutator $\{Q_{\alpha},\bar{Q}_{\dot{\beta}}\} \not=
0$. This allows chiral part noncommutativity $\{ \theta_{\alpha} , \theta_{\beta} \}
\not= 0$ or anti-chiral part noncommutativity $\{ \bar{\theta}_{\dot{\alpha}} , \bar{\theta}_{\dot{\beta}} \}
\not= 0$ but not both. Actually, this is possible only when we consider
Euclidean (4+0) or Atiyah-Ward (2+2) space-times \cite{Seiberg,Ketov_Sasaki}.

Other interesting twist are possible. For example
\begin{equation}
\mathcal{F}^{PQ} = \exp\left[ \frac{i}{2} \lambda^{\mu \alpha} ( P_{\mu} 
\otimes Q_{\alpha} - Q_{\alpha} \otimes P_{\mu}) \right]
\end{equation}
gives mixed noncommutativity between bosonic and fermionic coordinates
\begin{eqnarray}
& &\left[x^{\mu},x^{\nu} \right]_{\star} = \lambda^{\mu \alpha} 
\sigma^{\nu}_{\alpha \dot{\beta}} \bar{\theta}^{\dot{\beta}} - 
\lambda^{\nu \alpha} \sigma^{\mu}_{\alpha \dot{\beta}} 
\bar{\theta}^{\dot{\beta}}, \nonumber \\
& & \left[ x^{\mu},\theta^{\alpha}\right]_{\star} = i \lambda^{\mu 
\alpha} \not=0, \nonumber \\
& & \left\{ \theta^{\alpha}, \theta^{\beta} \right\}_{\star} = 0.
\end{eqnarray}
It should be confirmed too that this twist element indeed satisfy twist 
equation. Unlike the ordinary noncommutative case Eq.~(\ref{P_twist}) 
(pure bosonic case), it is less obvious whether this fermionic twist 
element satisfies the twist equation and Yang-Baxter equation because of 
bothersome sign flips from interchanging grassmann odd quantity.
But regarding $\lambda^{\mu \alpha} Q_{\alpha}$ as {\it bosonic}
quantity, the same procedure in preceding section 
can be applied to the twist equation.

We can consider more general setting such as
\begin{equation}
\mathcal{F} = \mathrm{exp}\left[ \frac{i}{2}\Theta^{\mu \nu} P_{\mu} \otimes 
P_{\nu} + \frac{i}{2} \lambda^{\mu \alpha} ( P_{\mu} \otimes Q_{\alpha} 
- Q_{\alpha} \otimes P_{\mu} ) - 
\frac{1}{2} C^{\alpha \beta} Q_{\alpha} \otimes Q_{\beta}  \right], \label{general_twist}
\end{equation}
where $\Theta^{\mu \nu} , C^{\alpha \beta}$ are 
some grassmann even constants and $ \lambda^{\mu \alpha}$ odd one. The fact that the all generators $P_{\mu} , Q_{\alpha}$
(anti)commute with each other allows one to calculate the twist equation more 
easily. In fact, all $ \Theta^{\mu \nu} P_{\mu} \otimes 
P_{\nu} , \  \lambda^{\mu \alpha}  P_{\mu} \otimes Q_{\alpha} 
, \ \lambda^{\mu \alpha} Q_{\alpha} \otimes P_{\mu} , \ C^{\alpha \beta} 
Q_{\alpha} \otimes Q_{\beta}$ are commute, then exponential will be factorized
\begin{equation}
\mathcal{F} = \mathcal{F}^{PP} \mathcal{F}^{QQ} \mathcal{F}^{PQ}.
\end{equation}

Notice that if each of this factorized sector satisfy the twsit equation 
separately, 
then we can see that all combined exponential Eq.~(\ref{general_twist}) also 
satisfy the twist equation. For example, if $P$-$P$ twist and $Q$-$Q$ 
twist satisfy the twist equation
\begin{eqnarray}
\left(\mathcal{F}^{PP}\right)_{12}
( \Delta_0 \otimes \mathrm{id}) \mathcal{F}^{PP}
 &=& \left( \mathcal{F}^{PP}\right)_{23} ( \mathrm{id} \otimes \Delta_0)
\mathcal{F}^{PP}, \label{PP_twist} \\
\!\!\! \!\! \!\! \!\!  \left( \mathcal{F}^{QQ}\right)_{12}
( \Delta_0 \otimes \mathrm{id})
\mathcal{F}^{QQ}
 &=& \left( \mathcal{F}^{QQ}\right)_{23} 
( \mathrm{id} \otimes \Delta_0)  \mathcal{F}^{QQ} \label{QQ_twist}
\end{eqnarray}
then
\begin{eqnarray}
& &\left( e^{\frac{i}{2} \Theta^{\mu \nu}  P_{\mu} 
\otimes P_{\nu} - \frac{1}{2} C^{\alpha \beta} Q_{\alpha} \otimes Q_{\beta} } \right)_{12} ( 
\Delta_0 \otimes \mathrm{id}) \left( e^{\frac{i}{2} \Theta^{\mu \nu}  P_{\mu} 
\otimes P_{\nu} - \frac{1}{2} C^{\alpha \beta} Q_{\alpha} \otimes Q_{\beta} } \right) \nonumber \\
&=& \left( \mathcal{F}^{PP}\right)_{12} \left(\mathcal{F}^{QQ} \right)_{12}
 \left(  e^{\frac{i}{2} \Theta^{\mu \nu}  \Delta_0(P_{\mu}) 
\otimes P_{\nu} - \frac{1}{2} C^{\alpha \beta} \Delta_0(Q_{\alpha}) \otimes Q_{\beta}}\right) 
\nonumber \\
&=& \left( \mathcal{F}^{PP} \right)_{12} 
\left( e^{ \frac{i}{2} \Theta^{\mu \alpha}  \Delta_0(P_{\mu}) 
\otimes P_{\nu}}\right)
\left(  \mathcal{F}^{QQ} \right)_{12}
\left( e^{- \frac{1}{2} C^{\alpha\beta} 
\Delta_0(Q_{\alpha}) \otimes Q_{\beta}  }\right) \nonumber \\
&=& \left(\mathcal{F}^{PP} \right)_{12} ( \Delta_0 \otimes \mathrm{id})
 \mathcal{F}^{PP} 
\left( \mathcal{F}^{QQ}\right)_{12}
 ( \Delta_0 \otimes \mathrm{id}) \mathcal{F}^{QQ} .
\end{eqnarray}
After using Eq.~(\ref{PP_twist}) and Eq.~(\ref{QQ_twist}) and factorization 
condition, we find this ($P$-$P$ + $Q$-$Q$) twist element actually satisfy the twist 
equaton. This is valid for $P$-$Q$ sector too and Yang-Baxter equation 
is easier to prove.

This twist element allows genelal noncommutative superspace relation
\begin{eqnarray}
& & \left[ x^{\mu}, x^{\nu} \right]_{\star} = i \Theta^{\mu \nu} + 
C^{\alpha \beta} 
\sigma^{\mu}_{\alpha \dot{\gamma}} \sigma^{\nu}_{\beta \dot{\delta}} 
\bar{\theta}^{\dot{\gamma}} \bar{\theta}^{\dot{\delta}} 
+
\lambda^{\mu \alpha} \sigma^{\nu}_{\alpha \dot{\beta}} 
\bar{\theta}^{\dot{\beta}} - \lambda^{\nu \alpha} \sigma^{\mu}_{\alpha 
\dot{\beta}} \bar{\theta}^{\dot{\beta}}, \nonumber \\
& & \left[ x^{\mu} , \theta^{\alpha} \right]_{\star} = i \lambda^{\mu 
\alpha} - i C^{\alpha \beta} \sigma^{\mu}_{\beta \dot{\gamma}} 
\bar{\theta}^{\dot{\gamma}}, \nonumber \\
& & \left\{ \theta^{\alpha}, \theta^{\beta} \right\}_{\star} = C^{\alpha 
\beta}, \label{mixedcommutaion}
\end{eqnarray}
and any other commutator involving anti-chiral part of fermionic 
coordinates is zero. 

Another type of non(anti)commutativity of superspace (called $\kappa$-deformation) is considered in Ref.\cite{Lukierski}.

\section{Consistency \label{consistency}}
Let us check the transformation rule of noncommutative parameters. These
show the consistency of transformation property and algebra.
Because merely the co-product of Lorentz generator $M_{\mu \nu}$ and
anti-supercharge $ \bar{Q}^{\dot{\alpha}}$ are deformed by twisting,
it is sufficient to show what these transform
the non(anti)commutative parameters into. 
\subsection{$P$-$P$ sector}
For the $P$-$P$ twisted sector, the transformation property of
noncommutative parameter $\Theta^{\mu \nu}$ was already shown in
Ref.\cite{Nishijima}. In this case, the twisted co-product of Lorentz generator
$M_{\mu \nu}$ is 
\begin{eqnarray}
\Delta_t^{PP} \left( M_{\mu \nu} \right) &=& M_{\mu \nu} \otimes
 \mathbf{1} + \mathbf{1} \otimes M_{\mu \nu} \nonumber \\
& & - \frac{1}{2} \Theta^{\rho \sigma} \left[ (\eta_{\rho \mu} P_{\nu} -
					\eta_{\rho \nu} P_{\mu}) \otimes
				       P_{\sigma} + P_{\rho} \otimes
				       (\eta_{\sigma \mu} - \eta_{\sigma
				       \nu} P_{\mu})\right].
\end{eqnarray}
The co-product of $\bar{Q}^{\dot{\alpha}}$ in this stage is not changed.
Then the action of twisted Lorentz generator to the twisted function
$f^t_{\rho \sigma} \equiv x_{\{ \rho} \star x_{ \sigma \}}
$\footnote{This corresponds to the function $f^{\rho \sigma} = x^{\rho}
x^{\sigma}$ in the commutative space. Detailed explanation is written in
Ref.\cite{Nishijima}.} is $M^t_{\mu \nu} f^t_{\rho \sigma} = m_{t} \circ
\left( \Delta_t \left( (M_{\mu \nu})(x_{ \{ \rho} \otimes x_{\sigma \} }
)\right)\right)$. 

We take $x^{ [ \rho} \otimes x^{\sigma ] }$ to 
see the noncommutative parameter $\Theta^{\rho \sigma}$ and find 
\begin{eqnarray}
m_t \circ \left( \frac{1}{i} \Delta_t^{PP}( M_{\mu \nu}) \left(
							   x^{\rho}
							   \otimes
							   x^{\sigma} -
							   x^{\sigma}
							   \otimes x^{\rho}
							  \right) \right) =
 M^t_{\mu \nu} ( \Theta^{\rho \sigma}) = 0.
\end{eqnarray}
Although the left-hand side of this equation looks like as tensor,
it transforms in the twisted Lorentz invariant way, {\it i.e.}
constant.

\subsection{$Q$-$Q$ sector}
For $Q$-$Q$ twisted sector, our co-product of Lorentz generator is
\begin{eqnarray}
\Delta_t^{QQ} (M_{\mu \nu}) &=& M_{\mu \nu} \otimes \mathbf{1} +
 \mathbf{1} \otimes M_{\mu \nu} \nonumber \\
& & + \frac{i}{2} \left( \sigma_{\mu \nu} \right)_{\alpha}^{\;\, \gamma}
 C^{\alpha \beta} \left( Q_{\beta} \otimes Q_{\gamma} + Q_{\gamma}
		   \otimes Q_{\beta} \right),
\end{eqnarray}
and anti-supercharge is
\begin{eqnarray}
\Delta_t^{QQ} (\bar{Q}^{\dot{\alpha}}) &=&
\bar{Q}^{\dot{\alpha}} \otimes \mathbf{1} +
\mathbf{1} \otimes \bar{Q}^{\dot{\alpha}} \nonumber \\
&+& C^{\gamma \delta} \epsilon^{\dot{\alpha} \dot{\beta}}
\{ (\sigma^\rho )_{\gamma \dot{\beta}} P_\rho \otimes Q_\delta
- Q_\gamma \otimes (\sigma^\rho )_{\delta  \dot{\beta}} P_\rho
\}.
\end{eqnarray}
Then, twisted actions on the function $h^{\alpha \beta} =
\theta^{\alpha} \theta^{\beta}$ are
\begin{eqnarray}
m_t \circ \left( \Delta_t^{QQ} (M_{\mu \nu}) \left( \theta^{\alpha}
					      \otimes \theta^{\beta}
						 \right)\right)
 = - i \left( \sigma_{\mu \nu} \right)_{\gamma}^{\;\, \alpha}
 \theta^{\gamma} \theta^{\beta} - i \left( \sigma_{\mu \nu}
				    \right)_{\gamma}^{\;\, \beta}
 \theta^{\alpha} \theta^{\gamma}, 
\end{eqnarray}
\begin{eqnarray}
m_t \circ \left( \Delta_t^{QQ} (\bar{Q}^{\dot{\alpha}}) \left( \theta^{\alpha}
					      \otimes \theta^{\beta}
						 \right)\right)
 = 0.
\end{eqnarray}

Twisted Lorentz transformation of non(anti)commutative parameter
$C^{\alpha \beta}$ can be calculated
\begin{eqnarray}
m_t \circ \left( \Delta_t^{QQ} (M_{\mu \nu}) \left(  \theta^{\alpha}
					      \otimes \theta^{\beta} +
					      \theta^{\beta} \otimes
					      \theta^{\alpha}
					     \right)\right) = M_{\mu
\nu}^t (C^{\alpha \beta}) = 0,
\end{eqnarray}
\begin{eqnarray}
(\bar{Q}^{\dot{\alpha}})^t(C^{\alpha \beta}) = 0.
\end{eqnarray}
So, non(anti)commutative parameter $C^{\alpha \beta}$ is also twisted Lorentz
invariant.

\subsection{$P$-$Q$ sector}
The co-products in this case are
\begin{eqnarray}
\Delta_t^{PQ} (M_{\mu \nu}) &=& M_{\mu \nu} \otimes \mathbf{1} +
 \mathbf{1} \otimes M_{\mu \nu} \nonumber \\
& & - \frac{1}{2} \lambda^{\rho \alpha} \left[ \left( \eta_{\mu \rho}
					       P_{\nu} - \eta_{\nu \rho}
					      P_{\mu} \right) \otimes
Q_{\alpha} - Q_{\alpha} \otimes \left( \eta_{\mu \rho} P_{\nu} -
				 \eta_{\nu \rho} P_{\mu} \right)
				       \right. \nonumber \\
& & \qquad \qquad \qquad - \left. P_{\rho} \otimes (\sigma_{\mu
 \nu})_{\alpha}^{\;\, \gamma} Q_{\gamma} + (\sigma_{\mu
 \nu})_{\alpha}^{\;\, \gamma} Q_{\gamma} \otimes P_{\rho} \right],
\end{eqnarray}
\begin{eqnarray}
\Delta_t^{PQ} (\bar{Q}^{\dot{\alpha}}) &=&
\bar{Q}^{\dot{\alpha}} \otimes \mathbf{1} +
\mathbf{1} \otimes \bar{Q}^{\dot{\alpha}} \nonumber \\
&+& \lambda^{\kappa \gamma} \epsilon ^{\dot{\alpha} \dot{\beta}}
( \sigma ^{\rho})_{\gamma \dot{\beta}} \left( 
 P_{\kappa} \otimes P_{\rho} - P_{\rho} \otimes P_{\kappa}
\right). 
\end{eqnarray}

So, same calculation shows
\begin{eqnarray}
m_t \circ \left( \frac{1}{i} \Delta_t (M_{\mu \nu}) (x^{\rho} \otimes
	   \theta^{\alpha} - \theta^{\alpha} \otimes x^{\rho} )\right) =
M_{\mu \nu}^t (\lambda^{\rho \alpha}) = 0,
\end{eqnarray}
\begin{eqnarray}
\bar{Q}^{\dot{\alpha}t}(\lambda^{\rho \alpha}) = 0.
\end{eqnarray}
Then $\lambda^{\mu \alpha}$ is twisted Loretnz invariant.
\subsection{Mixed sector}
Actually, in $P$-$P$ + $Q$-$Q$ + $P$-$Q$ mixed twist sector 
it results in only a linear combination of the three sectors:  

\begin{eqnarray}
\Delta_t^{Mix} (M_{\mu \nu}) &=& M_{\mu \nu} \otimes \mathbf{1} +
 \mathbf{1} \otimes M_{\mu \nu} \nonumber \\
& & - \frac{1}{2} \Theta^{\rho \sigma} \left[ (\eta_{\rho \mu} P_{\nu} -
					\eta_{\rho \nu} P_{\mu}) \otimes
				       P_{\sigma} + P_{\rho} \otimes
				       (\eta_{\sigma \mu} P_{\nu} - \eta_{\sigma
				       \nu} P_{\mu}) \right] \nonumber \\
& & + \frac{i}{2} \left( \sigma_{\mu \nu} \right)_{\alpha}^{\; \, \gamma}
 C^{\alpha \beta} \left( Q_{\beta} \otimes Q_{\gamma} + Q_{\gamma}
		   \otimes Q_{\beta} \right) \nonumber \\ 
& & - \frac{1}{2} \lambda^{\rho \alpha} \left[ \left( \eta_{\mu \rho}
					       P_{\nu} - \eta_{\nu \rho}
					      P_{\mu} \right) \otimes
Q_{\alpha} - Q_{\alpha} \otimes \left( \eta_{\mu \rho} P_{\nu} -
				 \eta_{\nu \rho} P_{\mu} \right)
				       \right] \nonumber \\
& & \qquad \qquad \qquad - \left[ P_{\rho} \otimes (\sigma_{\mu
 \nu})_{\alpha}^{\;\, \gamma} Q_{\gamma} + (\sigma_{\mu
 \nu})_{\alpha}^{\;\, \gamma} Q_{\gamma} \otimes P_{\rho} \right] ,
\end{eqnarray}
\begin{eqnarray}
\Delta_t^{Mix} (\bar{Q}^{\dot{\alpha}}) &=&
\bar{Q}^{\dot{\alpha}} \otimes \mathbf{1} +
\mathbf{1} \otimes \bar{Q}^{\dot{\alpha}} \nonumber \\
&+& C^{\gamma \delta} \epsilon^{\dot{\alpha} \dot{\beta}}
\{ (\sigma^\rho )_{\gamma \dot{\beta}} P_\rho \otimes Q_\delta
- Q_\gamma \otimes (\sigma^\rho )_{\delta  \dot{\beta}} P_\rho \} \nonumber \\
&+& \lambda^{\kappa \gamma} \epsilon ^{\dot{\alpha} \dot{\beta}}
( \sigma ^{\rho})_{\gamma \dot{\beta}} \left( 
 P_{\kappa} \otimes P_{\rho} - P_{\rho} \otimes P_{\kappa}
\right). 
\end{eqnarray}

All (anti)commutation relations are transformed as following.
\begin{eqnarray}
m_t \circ \left( \Delta_t (M_{\mu \nu}) (x^{\rho} \otimes
	   x^{\sigma} - x^{\sigma} \otimes x^{\rho} ) \right)
&=& i  \left[ \lambda^{\sigma \alpha}
(\sigma^{\rho})_{\alpha \dot{\alpha}}
(\bar{\sigma}_{\mu \nu} )^{\dot{\alpha}} _{\;\, \dot{\gamma}}
-
\lambda^{\rho \alpha}
(\sigma^{\sigma})_{\alpha \dot{\alpha}}
(\bar{\sigma}_{\mu \nu} )^{\dot{\alpha}} _{\;\, \dot{\gamma}}
\right] \bar{\theta}^{\dot{\gamma}},
\nonumber \\
m_t \circ \left( \Delta_t (M_{\mu \nu}) (x^{\rho} \otimes
	   \theta^{\alpha} - \theta^{\alpha} \otimes x^{\rho} ) \right) 
&=& C^{\gamma \alpha} ( \sigma_\rho )_{\gamma \dot{\kappa}}
( \bar{\sigma}_{\mu \nu} )^{\dot{\kappa}}_{\;\, \dot{\gamma}} \bar{\theta}^{\dot{\gamma}},
\nonumber \\
m_t \circ \left( \Delta_t (M_{\mu \nu}) (\theta^{\alpha} \otimes
	   \theta^{\beta} + \theta^{\beta} \otimes \theta^{\alpha} ) \right) 
&=& 0,
\end{eqnarray}
\begin{eqnarray}
m_t \circ \left( \Delta_t (\bar{Q}^{\dot{\alpha}}) (x^{\rho} \otimes
	   x^{\sigma} - x^{\sigma} \otimes x^{\rho} ) \right)
&=& -i C^{\gamma \delta} \epsilon^{\dot{\alpha} \dot{\beta}}  
\left[ (\sigma^\rho)_{\gamma \dot{\beta}} (\sigma^{\sigma} 
)_{\delta \dot{\delta}} \bar{\theta}^{\dot{\delta}}
- (\sigma^{\rho})_{\gamma \dot{\gamma}} \bar{\theta}^{\dot{\gamma}}
( \sigma^{\sigma}) _{\delta \dot{\beta}}
\right] 
\nonumber \\
& & -i \epsilon^{\dot{\alpha} \dot{\beta}}
\left[ \lambda^{\rho \gamma} (\sigma ^ \sigma)_{\gamma \dot{\beta}}
- \lambda^{\sigma \gamma} (\sigma ^ \rho)_{\gamma \dot{\beta}}
\right],
\nonumber \\
m_t \circ \left( \Delta_t (\bar{Q}^{\dot{\alpha}}) 
(x^{\rho} \otimes \theta^{\alpha} - \theta^{\alpha} \otimes x^{\rho})
 \right)
&=& -C^{\gamma \alpha} \epsilon^{\dot{\alpha} \dot{\beta}}
(\sigma ^ \rho)_{\gamma \dot{\beta}}, 
\nonumber \\
m_t \circ \left( \Delta_t (\bar{Q}^{\dot{\alpha}}) 
(\theta^{\alpha} \otimes
	   \theta^{\beta} + \theta^{\beta} \otimes \theta^{\alpha} ) \right) 
&=& 0.
\end{eqnarray}
These consist with the tansformation of Eq.~(\ref{mixedcommutaion}),
such that non(anti)commutative parameters are invariant,
while coordinates are transformed precisely.

\section{Extended non(anti)commutative (super)space\label{extended}}
Non(anti)commutativity of extended superspace $ \{ \theta^I_{\alpha} ,
\theta^J_{\beta} \} = C^{IJ}_{\alpha \beta} \not= 0$ has also been considered. 
Especially $\mathcal{N}=2$ non(anti)commutative superspace \cite{ferrara} and field theory 
on it was 
studied \cite{Ketov_Sasaki,Araki_Ito_Ohtsuka,Ivanov_Lechtenfeld,Ferrara_Ivanov}. But 
extended Poincar\'e SUSY algebra
\begin{eqnarray}
[P_{\mu},Q_{\alpha}^I] &=& 0 , \quad [M_{\mu \nu} , Q_{\alpha}^I] = i ( 
\sigma_{\mu \nu})_{\alpha}^{\beta} Q_{\beta}^I ,\nonumber \\
\{Q_{\alpha}^I, \bar{Q}_{\dot{\beta}}^J\} &=& 2 \sigma^{\mu}_{\alpha 
\dot{\beta}} P_{\mu} \delta^{IJ} , \quad \{ Q_{\alpha}^I, Q_{\beta}^J \} 
= \varepsilon_{\alpha \beta} Z^{IJ},
\end{eqnarray}
shows that it looks impossible to use $Q_{\alpha}^I$ generator to construct 
twist element because of their non-abelianity\footnote{But in some special setting, for example
$\mathcal{N}= 2$ singlet deformed superspace can admit Lorentz invariant
deformation of superspace. For more detail, see Ref.\cite{ferrara}}. Instead, we can use central charge generator $Z^{IJ}$ 
because central charge always commutes with any other generators. Now we 
set up
\begin{equation}
\mathcal{F} = \exp \left( \frac{i}{2} \Xi_{IJ} \ Z^I \otimes Z^J \right),
\end{equation}
$\Xi_{IJ}$ is some constant. This $Z$-$Z$ twist also satisfies the twist 
equation. The proof is the same as  $P$-$P$ twist by their bosonic 
abelian property.

Central charge coodinate formulation of supersymmetric
theory \cite{sohnius,Zheltukhin,Hewson} is needed when there are nonzero central charge in
the algebra. Central charge generator $ Z^I = \frac{\partial}{\partial z_I}
$ act on the central charge coordinate $z_J$ as
\begin{equation}
Z^I z_J = \delta^I_J
\end{equation}
then, one can easily compute the commutator of this coordinate
\begin{eqnarray}
m_t ( z_I \otimes z_J) &=& z_I \star z_J = m \circ e^{ -\frac{i}{2} \Xi_{KL} \ Z^K \otimes Z^L } (z_I \otimes z_J) \nonumber \\
&=& m \circ \left[ z_I \otimes z_J + \frac{i}{2} \Xi_{KL} \ \delta^K_I \otimes \delta^L_J \right] \nonumber \\
&=& z_I z_J + \frac{i}{2} \Xi_{IJ}.
\end{eqnarray}
Now $\Xi_{IJ}$ is antisymmetric with respect to $I,J$, we obtain
\begin{equation}
[z_I,z_J]_{\star} = i \,\Xi_{IJ} \not=0 .
\end{equation}
This is noncommutative relation among central charge coordinate. But
this situation is possible only when we consider extended $ \mathcal{N}
\ge 3 $ SUSY. The reason is that $\mathcal{N} = 1 $ SUSY does not contain central
charge and $ \mathcal{N} = 2 $ SUSY have only one central charge $
Z^{12} = - Z^{21} \equiv Z $ so it always commute.

\section{Conclusion\label{Conclusion}}
In this paper, we extend twisted deformed (Hopf) Poincar\'e algebra 
$\mathcal{U}_t \mathcal{(P)}$ 
based on Ref.\cite{Nishijima} to supersymmetric one $\mathcal{U}_t \mathcal{(SP)}$. Considering this 
symmetry as our fundamental symmetry, we can set up quantum field theory on 
non(anti)commutative space by twisted Lorentz invariant way. More practical work 
was done in Ref.\cite{Chaichian} for bosonic (non supersymmetric) sector. This approach may 
useful for recent developing supersymmetric gauge theory on 
non(anti)commutative superspace. We hope these reinterpretaiton of 
symmetry algebra helps us to consider quantum field theory on 
noncommutative space correctly and ameliorates some ambiguity of representation.
In our approach, it seems that $\mathcal{N}$=1 twisted SUSY remain completely unbroken,
because our twisting procedure doesn't change the original algebra.
Which situation is different from ordinary approach that breaks half of
SUSY and settle down to $\mathcal{N}$= 1/2 SUSY after turning on non(anti)
commutativity of superspace \cite{Seiberg}. It may give a new physical
insight on this reinterpreted formalism.
In this paper, we consider only $\mathcal{N}=1$ 
simplest non(anti)commutative space. It seems difficult to extend to 
$\mathcal{N} \ge 2$ case. But instead we can consider the theory on 
noncommutative central charge coordinate $z_I$. It is interesting to 
investigate the physical meaning of this noncommutativity. These 
subjects are the future research interest that relate this twisted 
algebraic approach.

\appendix
\section{Appendix}
Following Ref.\cite{Nishijima}, our super Poincar\'e algebraic relation and
representation of this algebra is {\it not} deformed from ordinary case
but product should be replaced by star product (twisted multiplication).
So, in case of commutative limit, we can easily recover the usual
representation and product between functions.
\subsection{Super Poincar\'e algebra}
Our notation of super Poincar\'e algebra is as follows.
\begin{eqnarray}
& & [ P_{\mu} , P_{\nu} ] = 0 ,\nonumber \\
& & [M_{\mu \nu} , M_{\rho \sigma}] = i \eta_{\nu \rho} M_{\mu \sigma} - i \eta_{\mu \rho} M_{\nu \sigma} - i \eta_{\nu \sigma} M_{\mu \rho} + i \eta_{\mu \sigma} M_{\nu \rho} ,\nonumber \\
& & [M_{\mu \nu} , P_{\rho}] = - i \eta_{\rho \mu} P_{\nu} + i \eta_{\rho \nu} P_{\mu} ,\nonumber \\
& & [P_{\mu} , Q^I_{\alpha}] = 0 , \quad [P_{\mu} , \bar{Q}^I_{\dot{\alpha}}] = 0 ,\nonumber \\
& & [M_{\mu \nu} , Q^I_{\alpha}] = i \left( \sigma_{\mu \nu} 
\right)_{\alpha}^{\;\, \beta} Q_{\beta}^I , \quad [M_{\mu \nu} , \bar{Q}^{I 
\dot{\alpha}}] = i \left( \bar{\sigma}_{\mu \nu}  
\right)^{\dot{\alpha}}_{\;\, \dot{\beta}} \bar{Q}^{I \dot{\beta}} ,\nonumber \\
& & \{ Q^I_{\alpha} , \bar{Q}^J_{\dot{\beta} }\} = 2 \sigma^{\mu}_{\alpha \dot{\beta}} P_{\mu} \delta^{IJ} ,\nonumber \\
& & \{ Q^I_{\alpha} , Q^J_{\beta} \} = \varepsilon_{\alpha \beta} Z^{IJ} , \quad \{ \bar{Q}^I_{\dot{\alpha}} , \bar{Q}^J_{\dot{\beta}}\} = \varepsilon_{\dot{\alpha} \dot{\beta}} Z^*_{IJ}.
\end{eqnarray}
Translation generator $P_{\mu}$, Lorentz generator $M_{\mu\nu}$ and
supercharge $Q_{\alpha}$ in $\mathcal{N}=1$ super Poincar\'e algebra can be represented as differential operator on superspace
\begin{eqnarray}
P_{\mu} &=&  i \partial_{\mu} , \nonumber \\
M_{\mu\nu} &=& i ( x_{\mu} \partial_{\nu} - x_{\nu} \partial_{\mu}) - i \theta^{\alpha} \left( \sigma_{\mu \nu} \right)_{\alpha}^{\;\, \beta} \frac{\partial}{\partial \theta^{\beta}} - i \bar{\theta}_{\dot{\alpha}} \left( \bar{\sigma}_{\mu \nu} \right)^{\dot{\alpha}}_{\;\, \dot{\beta}} \frac{\partial}{\partial \bar{\theta}_{\dot{\beta}}},\nonumber \\
Q_{\alpha} &=&  i \frac{\partial}{\partial \theta^{\alpha}} - \sigma^{\mu}_{\alpha \dot{\beta}} \bar{\theta}^{\dot{\beta}} \partial_{\mu},
\quad
\bar{Q}_{\dot{\alpha}} = -i \frac{\partial}{\partial \bar{\theta}^{\dot{\alpha}}}  + \theta^{\beta} \sigma^{\mu}_{\beta \dot{\alpha}} \partial_{\mu}. 
\end{eqnarray}
We take the convention that $\sigma_{\mu \nu} = 
- \frac{1}{4} (\sigma_{\mu} \bar{\sigma}_{\nu}
- \sigma_{\nu} \bar{\sigma}_{\mu} )$.

\end{document}